\documentclass[unnumsec,webpdf,contemporary,large]{oup-authoring-template}%

\usepackage{float}
\usepackage{booktabs}
\usepackage{pifont}
\usepackage{pdflscape}

\graphicspath{{Fig/}}

\theoremstyle{thmstyleone}%
\theoremstyle{thmstyletwo}%
\theoremstyle{thmstylethree}%

\begin{document}

\journaltitle{Journal Title Here}
\DOI{DOI HERE}
\copyrightyear{2024}
\pubyear{2024}
\access{Advance Access Publication Date: Day Month Year}
\appnotes{Paper}

\renewcommand{\thefigure}{\arabic{figure}}
\setcounter{figure}{0}

\firstpage{1}


\title[AntiFold]{AntiFold: Improved antibody structure-based design using inverse folding}

\author[1,†]{Magnus Haraldson Høie}
\author[2,†]{Alissa Hummer}
\author[2]{Tobias H. Olsen}
\author[2]{Broncio Aguilar-Sanjuan}
\author[1]{Morten Nielsen}
\author[2,$\ast$]{Charlotte M. Deane}

\authormark{Høie and Hummer et al.}

\address[1]{\orgdiv{Department of Health Technology, Section for Bioinformatics}, \orgname{Technical University of Denmark}}
\address[2]{\orgdiv{Department of Statistics}, \orgname{University of Oxford}, \country{United Kingdom}}

\corresp[†]{These authors contributed equally to this work. }
\corresp[$\ast$]{Corresponding author \href{email:deane@stats.ox.ac.uk}{deane@stats.ox.ac.uk}}

\received{Date}{0}{Year}
\revised{Date}{0}{Year}
\accepted{Date}{0}{Year}



\abstract{
\textbf{Summary:}
The design and optimization of antibodies requires an intricate balance across multiple properties. Protein inverse folding models, capable of generating diverse sequences folding into the same structure, are promising tools for maintaining structural integrity during antibody design. Here, we present AntiFold, an antibody-specific inverse folding model, fine-tuned from ESM-IF1 on solved and predicted antibody structures. AntiFold outperforms existing inverse folding tools on sequence recovery across complementarity-determining regions, with designed sequences showing high structural similarity to their solved counterpart. It additionally achieves stronger correlations when predicting antibody-antigen binding affinity in a zero-shot manner, while performance is augmented further when including antigen information. AntiFold assigns low probabilities to mutations that disrupt antigen binding, synergizing with protein language model residue probabilities, and demonstrates promise for guiding antibody optimization while retaining structure-related properties.
\\
\textbf{Availability and implementation:}
AntiFold is freely available under the BSD 3-Clause as a web server (\url{https://opig.stats.ox.ac.uk/webapps/antifold/}) and pip installable package at: \url{https://github.com/oxpig/AntiFold}.
}


\maketitle

\section{Introduction}

Antibodies are one of the largest classes of therapeutics, used to treat diseases including cancers, autoimmune conditions and viral infections \citep{Lu2020}. Therapeutic antibody design is complex, requiring the optimization of numerous properties related to efficacy, manufacturability and safety \citep{Rabia2018}.

Machine learning-based methods have shown promise in accelerating multiple steps in the antibody development pipeline \citep{Hummer2022}, by reducing liabilities such as immunogenicity \citep{Marks2021, Prihoda2022, Tennenhouse2023} and aggregation \citep{Makowski2023}, or rationally optimizing for desirable properties such as binding affinity and developability \citep{Makowski2022}. 

A guiding consideration in antibody optimization is selecting mutations that maintain the structure and therefore structure-mediated properties, such as stability and antigen binding mode. Protein inverse folding models are trained to predict sequence given structure \citep{Ingraham2019,Strokach2020,Anand2022,Jing2021,Hsu2022,proteinmpnnDauparas2022}, and can therefore be used to design novel sequences without altering the antibody backbone structure. Backbone-constrained design could enable the optimization of individual properties without disrupting others.

Antibodies demonstrate distinct structure and sequence properties as compared to general proteins \citep{Stanfield2014, regep_2017_cdr_loops} (Figure~\ref{fig:antifold-fig1}). While the framework (FR) regions are germline-encoded and relatively conserved, the complementarity-determining region (CDR) loops are hypervariable and form most of the antigen-binding contacts. Over two thirds of CDRH3 loops have distinct structures not found in other general protein structures \citep{regep_2017_cdr_loops}.

Mutiple tools for antibody inverse folding design have recently been released, including AbMPNN \citep{Dreyer2023} and IgMPNN \citep{igdesign_absic}, based on the ProteinMPNN architecture \citep{proteinmpnnDauparas2022}. Furthermore, recent work has shown the promise of protein language and inverse folding models for guiding antibody affinity maturation through the identification of high-fitness, structurally-constrained regions of the mutational landscape \citep{hie_plm, inverse_folding_antibody_evolution, outeiral2024}. However, the sequence recovery of existing tools for the CDR regions has been limited. Additionally, ProteinMPNN-based tools have features such as the occasional re-ordering of antibody chains or CDRH3 position 112 insertions, or the introduction of gaps into IMGT numbered antibodies, incompatible with antibody structures.

\begin{figure}
  \centering
  \includegraphics[width=0.47\textwidth]{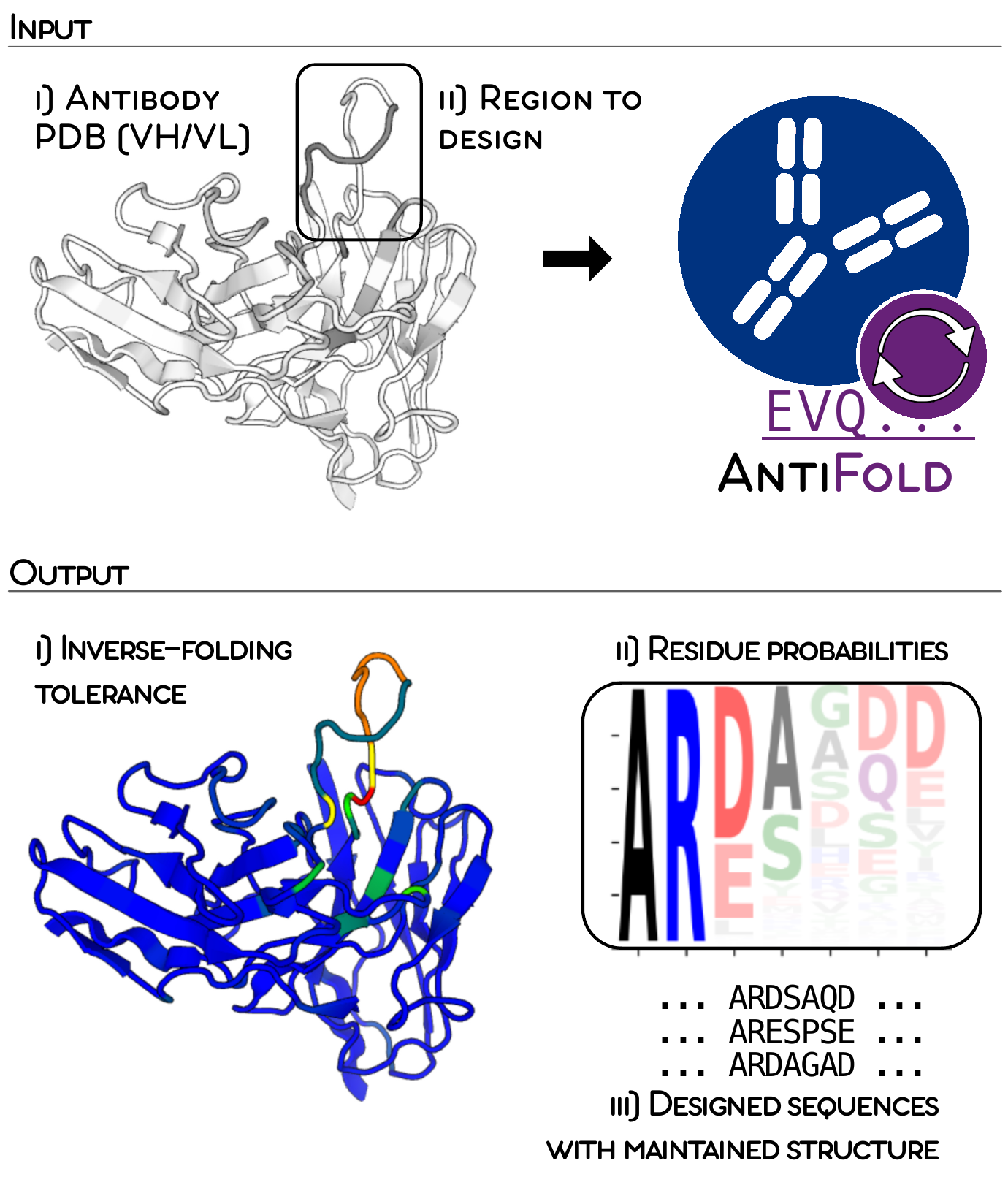}
  \caption{
  \textbf{Structure-constrained antibody design with AntiFold.} The user inputs an antibody variable domain PDB structure (heavy and light chain) and specifies an IMGT region to design. AntiFold outputs for each residue position in the PDB: i) structural tolerance to mutations without altering the backbone structure, ii) residue probabilities and iii) a number of designed sequences (default 10) for the selected region, predicted to maintain its structural fold. The diversity of the generated sequences may be controlled with a temperature parameter.}
  \label{fig:antifold-graphical}
\end{figure}

Here, we present AntiFold, an antibody-specific inverse folding model fine-tuned from ESM-IF1 \citep{Hsu2022}, which significantly improves upon CDR sequence recovery and zero-shot affinity prediction. AntiFold accepts an input solved or predicted antibody variable domain structure (Figure~\ref{fig:antifold-graphical}). For each residue position, the tool outputs the overall tolerance to mutations without altering the backbone structure (perplexity) and individual amino acid probabilities. Optionally, the user may specify regions to sample new sequences for, and a temperature parameter to control sequence diversity. Designed sequences show high structural similarity to the original structure when re-folded. The use of AntiFold in tandem with other property prediction tools could therefore guide antibody optimization campaigns by prioritizing experimental validation to a smaller search space.

\section{Methods}

\setlength{\textfloatsep}{12pt}

\subsection{Data}
AntiFold leverages ESM-IF1's pre-training on $>$12M structures by fine-tuning the model further on solved and predicted antibody structures. To enable a direct comparison with AbMPNN, we trained, validated and tested our model on the same data: 2,074 solved complexes from the Structural Antibody Database (SAbDab) \citep{Dunbar2014,Schneider2021}, and 147,458 structures of sequences from the Observed Antibody Space (OAS) paired database \citep{Kovaltsuk2018,Olsen2022} modeled with ABodyBuilder2 \citep{Abanades2023}. Each dataset was split using a 90\% concatenated CDR sequence identity cutoff (80/10/10 train/validation/test) \citep{Dreyer2023}. We did not directly compare against IgMPNN as the model weights and training data were not available.

\subsection{Training AntiFold}
During fine-tuning, we applied several strategies to improve performance on the validation set, specifically amino acid recovery (AAR) of the heavy chain CDR3 loop (CDRH3), which forms most of the antigen binding site interactions. CDRH3 AAR was improved through use of a span-masking scheme (36.4\%), additional masking of random residues (53.2\%), weighting masking towards CDR residues with a 3:1 weight (53.3\%), layer-wise learning-rate decay (54.4\%) and including predicted structures from OAS (58.4\%) (Tables~\ref{tab:fine-tuning_exp}-3). These augmentations were all included in the final AntiFold model. Training AntiFold without the ESM-IF1 pre-training yielded a CDRH3 AAR of 31.5\% on the validation set (Table \ref{tab:fine-tuning_no_pre-training}), highlighting the value of ESM-IF1's large pre-training dataset. For more details on these results and our fine-tuning strategy see the Supplementary Information.

\section{Results}

\subsection{Fine-tuning improves amino acid recovery on antibody sequences}

\begin{table*}[htbp]
\centering
\caption{\textbf{Summary of AntiFold performance on sequence design and binding affinity prediction.} For detailed methods and results, see the Supplementary Materials. Arrows indicate where higher/lower values are better: the best results are shown in bold. Evaluations from top to bottom; AbMPNN test-set amino acid recovery (Figure ~\ref{fig:antifold-fig2}), sequence design of CDR loops backbone RMSD (Figure~\ref{fig:supp_a3_RMSD}), \cite{warszawski_vhvl_dms} antibody-antigen deep mutational scan (Figure~\ref{fig:supp_a4_dms}), separation of \cite{hie_plm} affinity maturation-improved AbAg variants (fold-change \textgreater 1.25) (Figure~\ref{fig:antifold-fig4}).}
\label{tab:summary_table}
\begin{tabular}{@{}cccccc@{}}
\toprule
\multirow{2}{*}{\textbf{Evaluation}}                                                            & \multirow{2}{*}{\textbf{Dataset}}                                           & \multicolumn{4}{c}{\textbf{Model}}                                                     \\ \cmidrule(l){3-6} 
                                                                                                &                                                                             & \textit{ProteinMPNN} & \textit{ESM-IF1} & \textit{AbMPNN} & \textit{\textbf{AntiFold}} \\ \midrule
\begin{tabular}[c]{@{}c@{}}Amino acid recovery,\\ CDRH3\\ (\%, $\uparrow$)\end{tabular}                     & \begin{tabular}[c]{@{}c@{}}AbMPNN\\ test-set\end{tabular}                   & 35\%                 & 43\%             & 56\%            & \textbf{60\%}              \\ \midrule

\begin{tabular}[c]{@{}c@{}}Sequence design,\\ sampled CDR loops\\ (RMSD, $\downarrow$)\end{tabular} & \begin{tabular}[c]{@{}c@{}}AbMPNN\\ test-set\end{tabular}                   & 1.03                    & 1.01             & 0.98            & \textbf{0.95}              \\ \midrule

\begin{tabular}[c]{@{}c@{}}AbAg\\ binding affinity\\ ($S_r$, $\uparrow$)\end{tabular}                          & \begin{tabular}[c]{@{}c@{}}Warszawski,\\ anti-lysozyme\\ DMS\end{tabular}   & 0.30                 & 0.32             & 0.33            & \textbf{0.42}              \\ \midrule

\begin{tabular}[c]{@{}c@{}}AbAg\\improved variants\\ (Rank \%, $\uparrow$)\end{tabular}           & \begin{tabular}[c]{@{}c@{}}Hie, 7x Ab\\ affinity-\\ maturation\end{tabular} & 73\%           & 57\%             & 55\%            & \textbf{80\%}              \\ \bottomrule

\end{tabular}
\end{table*}

AntiFold demonstrated a substantial improvement in AAR on the experimental structure test set as compared to the original ESM-IF1 model (60 vs 43\% AAR for CDRH3, p $<$ 0.005; Table~\ref{tab:summary_table}, Figure~\ref{fig:antifold-fig2}. AntiFold also outperformed AbMPNN for CDRH3 (60\% vs 56\%), all remaining CDR regions (75-84\% vs 63-76\%), and most FR regions (87-94\% vs 85-89\%, Figure ~\ref{fig:supp_a1_FWs}).

We confirmed that AntiFold can be accurately applied to modelled structure inputs by testing it on the test set structures predicted with ABodyBuilder2. AntiFold achieved similar AAR for solved and predicted structures \mbox{($\Delta$AAR c-0.5\%)}, unlike AbMPNN which performed slightly worse on experimental structures ($\Delta$AAR -2.7\%, Figure~\ref{fig:antifold-fig2}C). AntiFold also maintained performance when applied to an antibody structure predicted with AlphaFold (PDB 7M3N, Table~\ref{tab:sanity1} \citep{colabfold}).

\subsection{Predicted sequences show high structural agreement with experimental structures}
To assess whether mutations suggested by AntiFold preserve the backbone structure, we sampled and refolded CDR sequences for a set of high-quality structures. We identified 56 antibody structures in the test set that were solved using X-ray crystallography with a resolution below 2.5 Å. Next, we sampled 20 sequences for each antibody using ProteinMPNN, ESM-IF1, AbMPNN and AntiFold using a residue sampling temperature of 0.20 (for more details, see Supplementary Methods).

We modeled these sequences using ABodyBuilder2, aligned them with the FR backbone of their experimentally solved counterpart, then calculated the root-mean-square deviation (RMSD) over the CDR residues (for more details, see Supplementary Methods). As a baseline, we modeled the true sequences with ABodyBuilder2 (native, Figure~\ref{fig:supp_a3_RMSD}). AntiFold generated sequences with high structural similarity to the original backbone, with a mean CDR region RMSD of 0.95, versus AbMPNN 0.98, ESM-IF1 1.01, ProteinMPNN 1.03 and native RMSD of 0.63.

\subsection{Inverse folding predicts antibody-antigen binding affinity}

We assessed the ability of AntiFold and other inverse folding models to predict antibody-antigen binding affinity by applying them to a deep mutational scan of an anti-lysozyme antibody \citep{warszawski_vhvl_dms}. We calculated the log-likelihoods of the 2209 variable domain variants of Fab D44.1 (PDB 1MLC) (Figure~\ref{fig:antifold-fig3}). As a sequence-only baseline, we included the sequence-based ESM-2 model (650M parameters). AntiFold significantly outperformed the other models with a Spearman's rank correlation of 0.418, versus ESM-IF1 (0.334), AbMPNN (0.322), ProteinMPNN (0.301) and ESM-2 (0.264), assessed by Mann-Whitney one-tailed U tests.

Next, we investigated whether adding the antigen chain(s) when calculating inverse folding probabilities would improve prediction of the experimental binding affinity (Figures~\ref{fig:supp_a7_dms_abag}, \ref{fig:supp_a8_dms_abag}). With this additional context, AntiFold's performance improves in the case of all CDR regions, in particular the CDR2 ($S_{r}$ 42.7 to 47.3\%, p $<$ 0.005) and CDR3 ($S_{r}$ 29.8 to 32.0\%, p $<$ 0.005), but not for framework residues. Contrary to this, ProteinMPNN and AbMPNN lose performance for most CDR regions when including the antigen chain, while performance on framework residues is largely unchanged.

To further explore the ability of AntiFold to guide antibody design by predicting antibody-antigen binding affinity, we applied the model to 124 variants across 7 antibodies generated in protein language model-guided affinity maturation experiments \citep{hie_plm}.

We first rank-normalized across all possible single amino acid variant scores for each antibody and each model. To assess the rankings of the 124 experimentally measured variants, we rank-normalized again across this set. Using the experimental binding affinity values, variants were separated into lower (fold-change $<$ 0.75), maintained (0.75-1.25) and improved ($>$ 1.25) binding affinity groups (Figure~\ref{fig:antifold-fig4}).

AntiFold achieved significantly improved separation of these groups, scoring the improved variants with a median rank score of 80\% (p $<$ 0.005) versus 73 \% for ProteinMPNN, 57 \% for ESM-IF1 and 55 \% for AbMPNN.

\subsection{Availability and Implementation}
AntiFold can be accessed as a web server or freely downloaded as a pip-installable package.
\begin{itemize}
    \item Web server: \url{https://opig.stats.ox.ac.uk/webapps/antifold/}
    \item Code repository: \url{https://github.com/oxpig/AntiFold}
\end{itemize}

\section{Discussion}
AntiFold, fine-tuned from the general protein inverse folding ESM-IF1, achieves state-of-the-art performance on antibody sequence recovery and refolding of designed sequences. The use of pre-trained weights substantially improves AntiFold's performance versus training from scratch, while including predicted structures and weighting masking towards individual CDR residues improved performance further.

AntiFold's inverse folding probabilities correlate with antibody-antigen binding affinity across multiple independent experiments. Including antigen information augments performance further, in the case of CDR residues close to the antigen binding site. These probabilities also synergize with protein language model-suggested variants, indicating the models learn orthogonal information. We find this performance is primarily driven by the de-selection of loss-of-binding variants. Consistent with previous results \citep{hie_plm, inverse_folding_antibody_evolution}, our findings indicate that AntiFold identifies structurally-constrained, high fitness regions of the mutational landscape, likely to preserve binding.

\section{Competing interests}
No competing interest is declared.

\section{Funding}
This work was supported by the Sino-Danish Center [2021, awarded to M.H.H], the Medical Research Council [grant number: MR/N013468/1 awarded to A.M.H], the Engineering and Physical Sciences Research Council [grant number: EP/S024093/1 awarded to T.H.O.] and GlaxoSmithKline plc.




\bibliographystyle{abbrvnat}
\bibliography{reference}


\clearpage
\newpage

\section{Supplementary information}
\setcounter{page}{1}
\renewcommand{\thetable}{S\arabic{table}}
\renewcommand{\thefigure}{S\arabic{figure}}
\setcounter{table}{0}
\setcounter{figure}{0}
\setcounter{section}{0}

\subsection{Supplementary Methods}

\subsubsection{Experimental antibody structures from SAbDab}
The AbMPNN dataset contains 2,074 structures of antibodies in complex with a protein antigen, after filtering for redundancy and experimental resolution $<$5 Å \citep{Dreyer2023}. We obtained structures of the corresponding variable fragment domains (Figure~\ref{fig:antifold-fig1}), numbered with the IMGT antibody numbering scheme \citep{LeFranc2003-IMGT} from SAbDab \citep{Dunbar2014,Schneider2021}. We modeled structures of the validation and test set using ABodyBuilder2 \citep{Abanades2023} to evaluate AntiFold performance on predicted structures. One and three structures were removed from the validation and test datasets, respectively, as these could not be modeled with ABodyBuilder2 \citep{Abanades2023}.

\subsubsection{Predicted antibody structures from ABodyBuilder2}
The structures of 148,832 paired antibody sequences from OAS \citep{Kovaltsuk2018,Olsen2022} modeled using ABodyBuilder2 were released as part of ABodyBuilder2 \citep{Abanades2023}. Filtering out structures with identical concatenated CDRs, as in AbMPNN \citep{Dreyer2023}, resulted in a dataset of 147,458 structures.

\subsubsection{Fine-tuning strategy}
We trained AntiFold by fine-tuning the ESM-IF1 inverse folding architecture \citep{Hsu2022} (Figure~\ref{fig:antifold-fig1}) on antibody structures. The inverse folding problem can be formalized as learning the conditional probability distribution, $p(Y|X)$, of the protein sequence, $Y$, consisting of amino acids $(y_1, \ldots, y_i, \ldots, y_n)$, given the structure, $X$, with spatial coordinates of the backbone atoms (N, C$_{\alpha}$ and C) $(x_1, \ldots, x_i, \ldots,x_{3n})$ (Equation (1)) \citep{Hsu2022}.

\begin{equation}
p(Y|X) = \prod_{i=1}^{n} p(y_i | y_{i-1}, \ldots, y_1; X)
\label{eq:inverse_folding}
\end{equation}

The ESM-IF1 architecture consists of 4 Graph Neural Network Geometric Vector Perceptron (GVP-GNN) layers \citep{Jing2021}, 8 generic Transformer \citep{Vaswani2017} encoder layers and 8 decoder layers \citep{Hsu2022}. The architecture is invariant to rotation and translation of the input coordinates.

The ESM-IF1 model is trained only on single chain structures. In order to represent complexes of antibody heavy and light chains, we concatenated the backbone coordinates of the light chain to the end of the heavy chain, with a 10 position padding of “gap” tokens, represented as missing coordinates in the input structure.

\subsubsection{Fine-tuning parameter evaluation}
We evaluated the effect of the parameters described below on model performance, as applied to the validation dataset.

\paragraph{\textbf{Layer-wise learning rate decay}}\mbox{} \\
We decayed the learning rate for each previous layer in the ESM-IF1 architecture by an alpha factor:

\begin{equation}
LR_i = LR \times \alpha^i
\label{eq:layer_decay}
\end{equation}

where $i$ ranges from zero to the number of layers in the model (20), and alpha is set to 0.85.

\paragraph{\textbf{Masking}}\mbox{} \\
We masked portions of the input antibody structure for model training and calculated loss over model predictions for the masked positions. The coordinates of masked positions were hidden for input to the model.

We evaluated three different masking schemes:
\begin{itemize}
    \item Shotgun masking: individual positions for masking were randomly selected 
    \item Span masking: spans (consecutive stretches of positions) were masked by randomly selecting starting positions and sampling the span length from a geometric distribution where p = 0.05, with a maximum span length of 30 positions
    \item Shotgun plus span masking: 7.5\% of the structure was first masked using span masking and a further 7.5\% was subsequently masked using the shotgun approach
\end{itemize}

Antibody sequence/structure can be separated into FR and CDR regions (Figure~\ref{fig:antifold-fig1}A), with the former being more conserved and typically easier to predict. As our model loss is calculated over masked positions, we explored whether performance could be improved by biasing the selection of masked positions towards CDR residues (IMGT-weighted masking). There are more than 2.5 times as many FR as CDR positions in the sequence. For shotgun masking, we implemented a 3:1 weighting for the selection of CDR vs FR positions. For span masking, we biased selection to be low (weight = 1) for most FR positions, high (weight = 3) for most CDR positions, and medium (weight = 2) for FR positions immediately preceding CDRs as well as CDR positions immediately preceding FRs.

\paragraph{\textbf{Gaussian noise}}\mbox{} \\
For predicted structures, we add noise to the backbone (N, C$_{\alpha}$ and C) 3D-coordinates, sampled from a Gaussian distribution with a scale of 0.1 Å, following the approach taken in ESM-IF1 \citep{Hsu2022}.

\subsubsection{Early stopping}
Model training was stopped when validation loss did not decrease after 10 epochs. The model with the lowest validation loss was carried forward.

\subsubsection{Model performance evaluation}
Amino acid recovery (AAR) is calculated as the percent of positions for which the top predicted amino acid is identical to the observed amino acid in the PDB.

\begin{equation}
\text{AAR} = \left( \frac{\sum_{i=1}^{N} \mathbf{1}[\text{argmax}(probs(i)) = \text{obs. AA}(i)]}{N} \right) \times 100\%
\end{equation}

where N is the total number of positions, and argmax(probabilities(i)) is a function that selects the amino acid with the highest probability at position i.

Model output probabilities are given by:

\begin{equation}
logits = raw\ model\ outputs
\label{eq:logits}
\end{equation}

\begin{equation}
probabilities(i) = \frac{e^{\text{logits}(i)}}{\sum_{j=1}^{20} e^{\text{logits}(j)}}
\label{eq:probabilities}
\end{equation}

Perplexity for each position is given by:

\begin{equation}
perplexities = 2^{-\sum_{i=1}^{20}probabilities(i) \times log_2(probabilities(i))}
\label{eq:perplexities}
\end{equation}

During sequence sampling, we sampled residues for each position in the CDRs proportional to their probability, using a temperature of 0.20. Here, a value of 1.00 represents sampling directly according to the model output distribution, while a value of $\sim$0.00 acts like an argmax function, ensuring only the top probability amino acid for each position is selected. We used the same method as ProteinMPNN \citep{proteinmpnnDauparas2022} of applying temperature directly to the logits before converting to probabilities:

\begin{equation}
scaled\ logits = \frac{logits}{t}
\label{eq:scaled_logits}
\end{equation}

ProteinMPNN \citep{proteinmpnnDauparas2022} and AbMPNN \citep{Dreyer2023} were run with default settings and the flags --conditional\_probs\_only, --sampling\_temp 0.20, --num\_seq\_per\_target 20 and --seed 37. Sampled sequences were then predicted with ABodyBuilder2 \citep{Abanades2023} at default settings. We corrected for ProteinMPNN reordered chains, reversal of insertions in IMGT positions 112 and invalid gaps.

We calculated RMSD using Pymol’s rms\_cur method \citep{PyMOL} between the solved and predicted backbone (N, C$_{\alpha}$, and C atoms) for each region, after aligning on the framework.

\subsubsection{Binding affinity prediction}
Inverse folding log-likelihoods were predicted for antibody variants in the \cite{warszawski_vhvl_dms} deep mutational scan for each of the inverse folding models by inputting the PDB 1MLC, heavy and light variable domains (IMGT positions 1-128). For ESM-2 (750M), we extracted and generated log-likelihoods for the corresponding antibody sequence. Experimental scores were mapped to a $log_{2}$ fold-change and correlated with inverse folding scores directly using scipy.stats.spearmanr \citep{2020SciPy-NMeth}.

Structures of antibody variants in the \cite{hie_plm} study were identified by searching the PDB for the extracted antibody sequence and selecting the highest sequence identity match. The X-ray structure with the highest resolutionw as selected.

\subsubsection{Rank normalization}
When assessing model ranking of improved amino acid variants, we first first rank-normalized all single amino variant scores ($N = L \times 20$) for each antibody separately. Next, we selected the 124 experimentally measured variants ($N=124$) and calculated their ranks using the same formula.

Rank normalization of scores was calculated as
$$
\text{Normalized Rank} = \frac{\text{Rank} - 1}{N - 1}
$$

Where Rank is the variant's score rank and N is the total number of variants.

\subsubsection{Bootstrapping}
For bootstrapping, we resampled with replacement 1000 times, with the bootstrapped values used to calculate means and confidence intervals.

\subsubsection{Statistical tests}
All reported p-values were calculated using the Mann-Whitney one-tailed U test unless otherwise stated.

\clearpage
\subsection{Supplementary results}

\subsubsection{Fine-tuning strategy}
Fine-tuning from a general protein inverse folding model enabled us to benefit from existing knowledge learned by ESM-IF1, which was trained on millions of structures. We explored the effect of multiple parameters on fine-tuning ESM-IF1 on antibody structures.

\begin{figure*}[h!]
  \centering
  \includegraphics[width=0.75\textwidth]{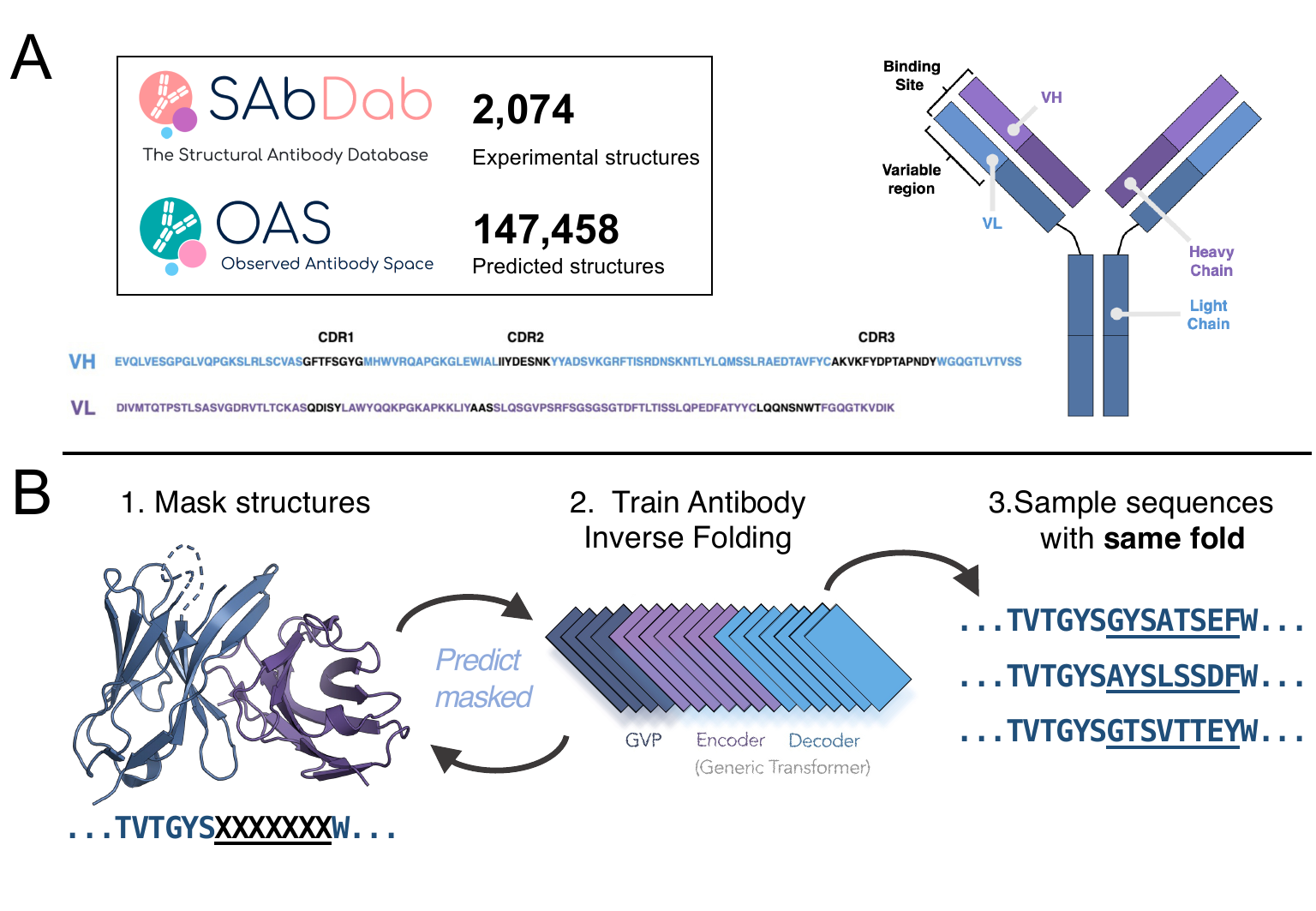}
  \caption{
  \textbf{Overview of the AntiFold training strategy.}
  (A) AntiFold was trained and evaluated on solved antibody structures from SAbDab \citep{Dunbar2014,Schneider2021} and structures of antibody sequences from OAS \citep{Kovaltsuk2018,Olsen2022} modeled with ABodyBuilder2 \citep{Abanades2023}. Antibodies consist of heavy (blue) and light (purple) chains. Target binding is primarily mediated by complementarity-determining regions (CDRs) in the variable region. Examples of heavy (VH) and light (VL) variable domain sequences are shown. (B) AntiFold is initialized with weights from ESM-IF1 \citep{Hsu2022}, then fine-tuned on antibody variable domain structures. AntiFold can generate diverse sequences maintaining the fold of the input structure. Figure adapted from  \citep{Olsen2022,Hsu2022}. Structure from PDB 3W2D \citep{pdb_3w2d}.}
  \label{fig:antifold-fig1}
\end{figure*}

\begin{figure*}[hbp]
  \centering
  \includegraphics[width=0.75\textwidth]{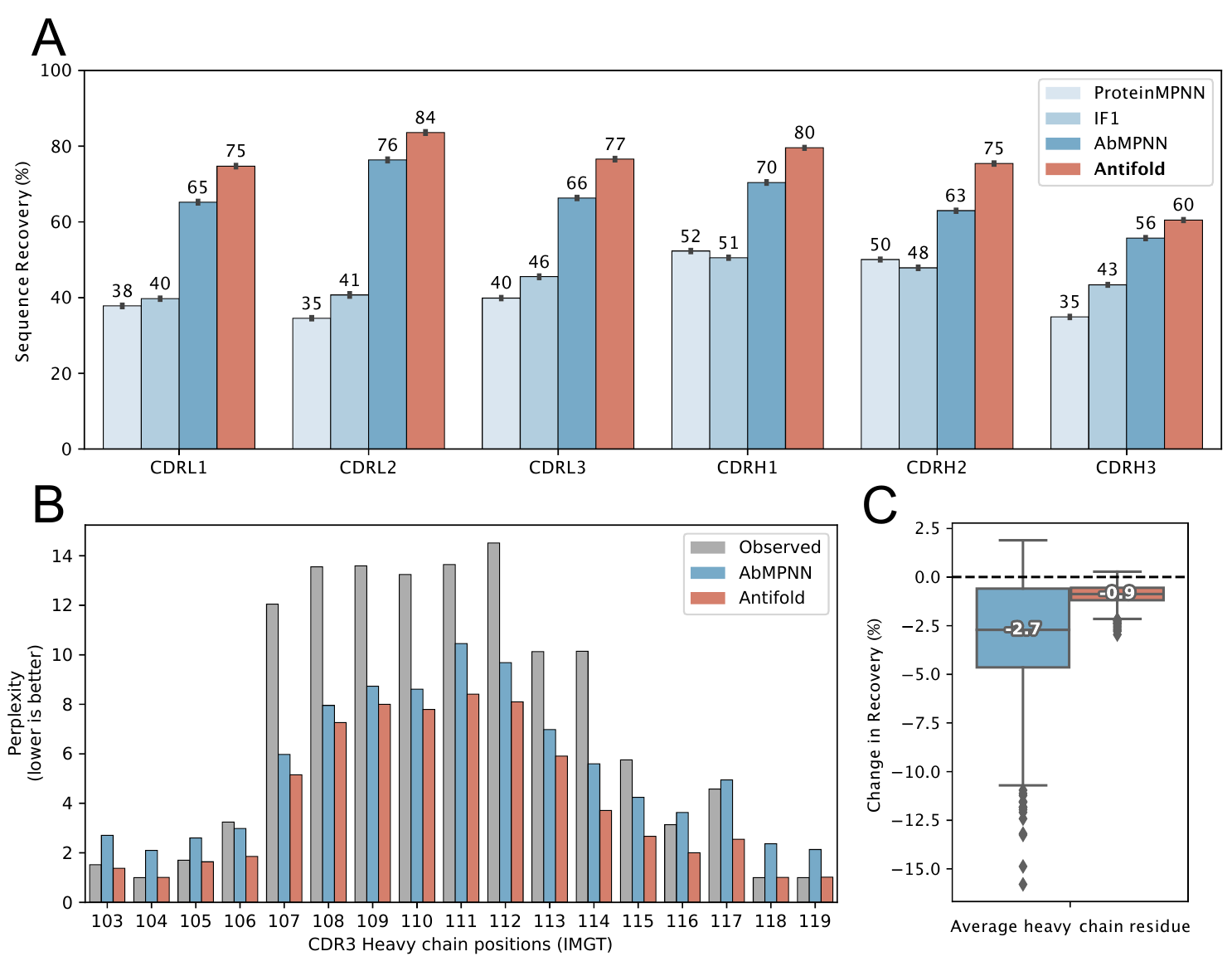}
  \caption{
  \textbf{AntiFold demonstrates improved amino acid recovery and perplexity.}
  (A) Mean amino acid recovery (AAR) across CDRs for antibody heavy and light chains. Error bars indicate 95\% confidence intervals after bootstrapping 100 times with replacement. (B) Perplexity across the CDRH3 loop. (C) Percent change in AAR when swapping predicted structures for experimental structures (same sequences). For details on these performance evaluations, see Supplementary Methods.}
  \label{fig:antifold-fig2}
\end{figure*}

\begin{figure*}[htbp]

  \centering
  \includegraphics[width=0.75\textwidth]{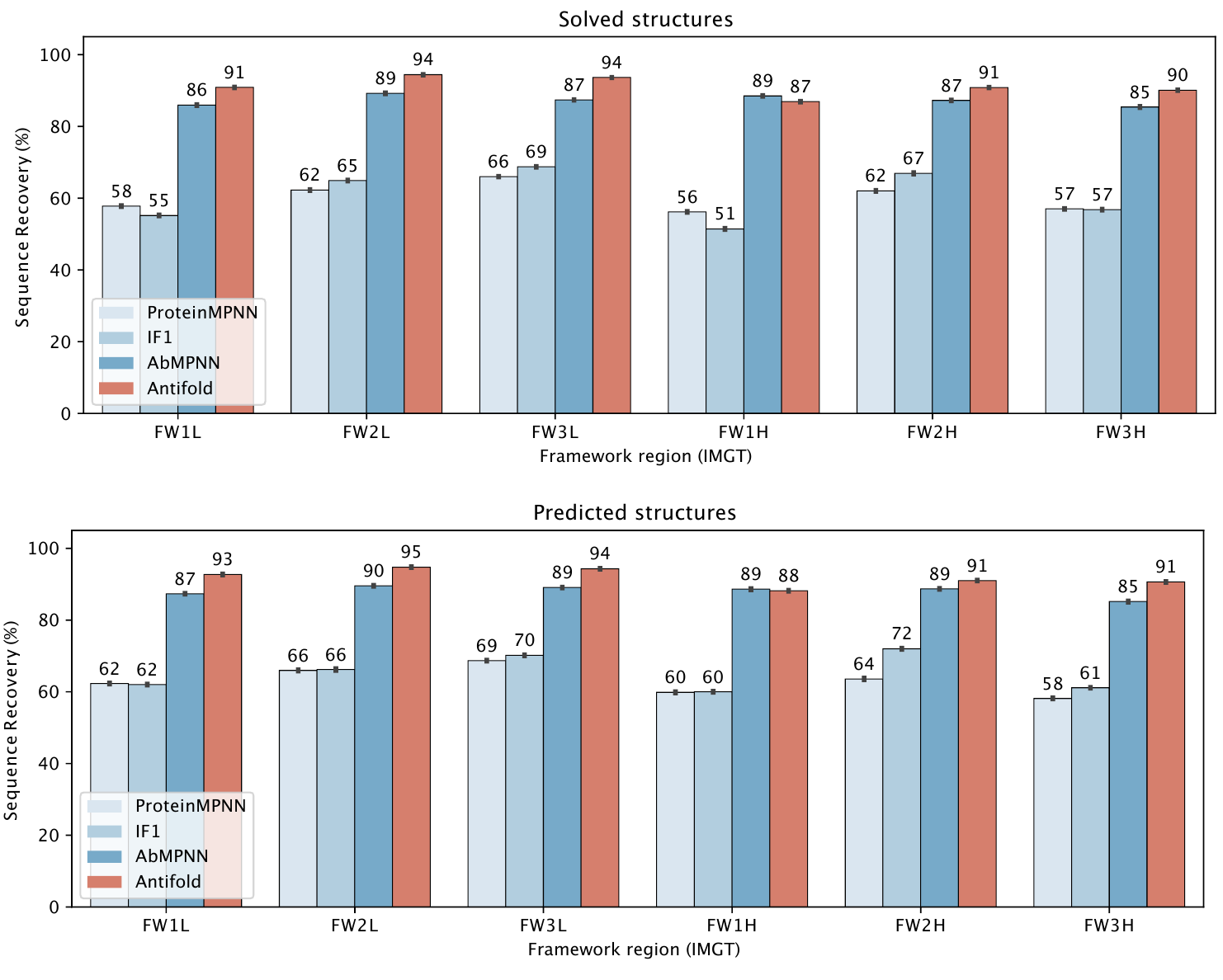}
  \caption{
  \textbf{Framework sequence recovery.}
  Framework amino acid sequence recovery, for solved (top) and predicted (bottom) structures in the test set.}
    \label{fig:supp_a1_FWs}
\end{figure*}

When fine-tuning on a new task or domain, there is a risk of “catastrophically forgetting” previously learned knowledge. We therefore applied a strategy of layer-wise learning rate decay, successfully used to fine-tune BERT models \citep{bert_finetuning}. We evaluated exponentially decaying the learning rate from the last to the first layer, preserving the weights of earlier parts of the model during training (see Figure~\ref{fig:antifold-fig1} and Methods). Layer-wise learning rate decay did not further improve sequence recovery (Supplementary Table A1-3), but we retained it for subsequent training to reduce the risk of overfitting and to maintain generalization towards untested properties.

We also investigated different masking schemes in training. Shotgun masking hides the coordinates of randomly selected single positions, while span masking is applied to a consecutive stretch of positions. As FR and CDR regions in the antibody structure have different levels of variability, we tested biasing the selection of masked positions towards the more variable CDR residues (3x weight, IMGT-weighted masking). In total, 15\% of the backbone residues were masked during training (for more details on the masking parameters, see Supplementary Methods). As previously reported \citep{Hsu2022}, we found stronger performance for shotgun than span masking on test structures. However, span masking improved CDR sequence recovery for test cases with masked CDR loops, a realistic design use case (Supplementary Table A1-3). IMGT-weighted masking further improved performance on CDR loops, while only slightly reducing sequence recovery on FR regions (Supplementary Table A1-2).

To improve performance by training on more diverse antibodies, we included a large dataset of 147,458 predicted structures from OAS \citep{OAS} modelled with ABodyBuilder2 \citep{Abanades2023}, in our fine-tuning strategy. We tested the effects of adding Gaussian noise at a scale of 0.1 Å to the modeled protein backbone, previously found to improve performance \citep{Hsu2022,proteinmpnnDauparas2022}. We found no substantial effect, but have included it in our final model for robustness towards minor variations in input structures (Supplementary Table A3).

Based on these results, we chose to train the final AntiFold model with IMGT-weighted shotgun and span masking, layer-wise learning rate decay and added Gaussian noise on predicted structures. We note that these augmentations, along with the use of the larger pre-trained ESM-IF1 architecture (142M parameters) instead of ProteinMPNN (1.7M parameters), comprise the main differences with AbMPNN. We split the training of AntiFold into two phases. First we fine-tuned ESM-IF1 on one pass of the training dataset of predicted structures from OAS. Next we fine-tuned the model on the solved training dataset, stopping training when there was no further improvement in validation loss for 10 epochs. This model, termed AntiFold, was used for all subsequent analysis.

\subsubsection{Observed variability (perplexity)}
We calculated the observed perplexity, a measure reflecting the number of amino acids observed in true antibody sequences, for each CDRH3 position in the test-set. Here, a value of 1 means only one amino acid is ever observed, while 20 represents all twenty amino acids being observed at equal probability. AntiFold constrains the observed perplexity ($\sim$10-14 observed amino acids, IMGT positions 107-114) down to 4-8 amino acids which are predicted to preserve the backbone structure of the loop, versus $\sim$6-10 for AbMPNN (Figure~\ref{fig:antifold-fig2}C). This, combined with the improved sequence recovery, reflects AntiFold's greater confidence and accuracy in designing structurally-maintained antibody sequences.

\subsubsection{Performance by CDRH3 length}
AntiFold's recovery performance is lower for antibodies with longer CDRH3 loops (Figure~\ref{fig:supp_a5_cdrlen}), with a median AAR of 71\% for shorter loops (6-9 residues) and 48\% for longer loops (16-28 residues).

\subsubsection{De-selection of lower binding affinity variants}
Applying AntiFold to an anti-lysozyme deep mutational scan \citep{warszawski_vhvl_dms}, we find that AntiFold effectively separated antibody variants with lower ($log_{2}$ fold-change $<$ 0) versus improved ($log_{2}$ fold-change $>$ 0) binding affinity (Fig~\ref{fig:supp_a4_dms}), likely by identifying structurally disruptive mutations which are also likely to disrupt antigen binding. For example, choosing a score threshold where $\sim$50\% of variants are discarded, AntiFold de-selects $\sim$40\% of lower while maintaining $>$95\% of improved binding affinity variants (Figure~\ref{fig:supp_a4_dms}B, score threshold -11).

\subsubsection{AntiFold CDRH3 substitution matrix}
We investigated AntiFold's trends in substituting residues in the test set CDRH3 positions, by calculating a 20x20 matrix of AntiFold's median inverse folding probabilities. Antifold CDRH3 log probabilities loosely resemble BLOSUM62 substitution trends, while distinctly disfavoring mutations from alanine, or to proline and cysteine (Fig~\ref{fig:supp_a6_IFmatrix}). 

\clearpage
\section{Supplementary Tables and Figures}

\setlength{\textfloatsep}{12pt}


\setlength{\textfloatsep}{12pt}

\begin{figure*}[htbp]

  \centering
  \includegraphics[width=0.75\textwidth]{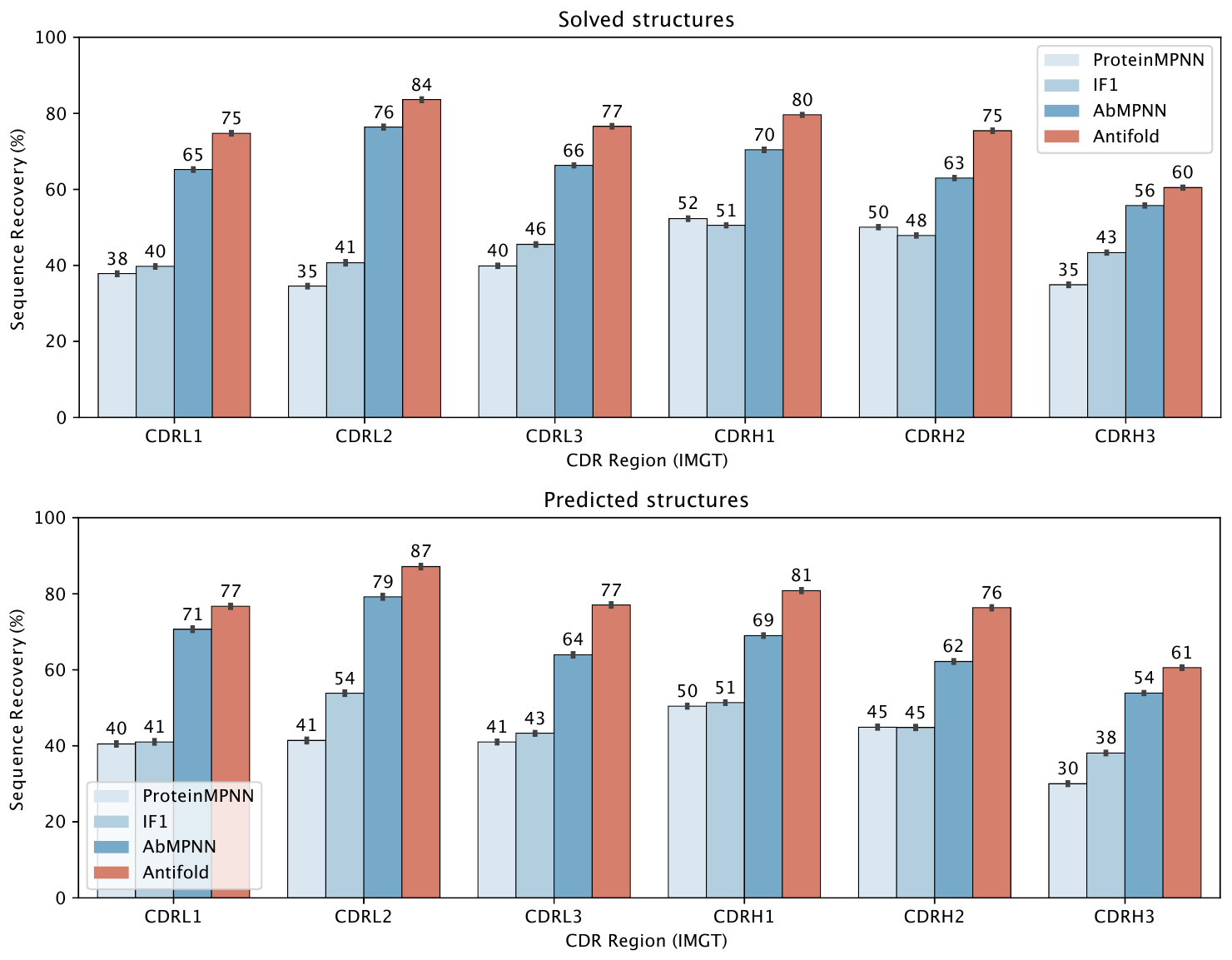}
  \caption{
    \textbf{CDR sequence recovery.}
  Complementarity determining region (CDR) amino acid sequence recovery for solved (top) and predicted (bottom) structures in the test set.}
    \label{fig:supp_a2_CDRs}
\end{figure*}

\begin{figure*}[htbp]
  \centering
  \includegraphics[width=0.75\textwidth]{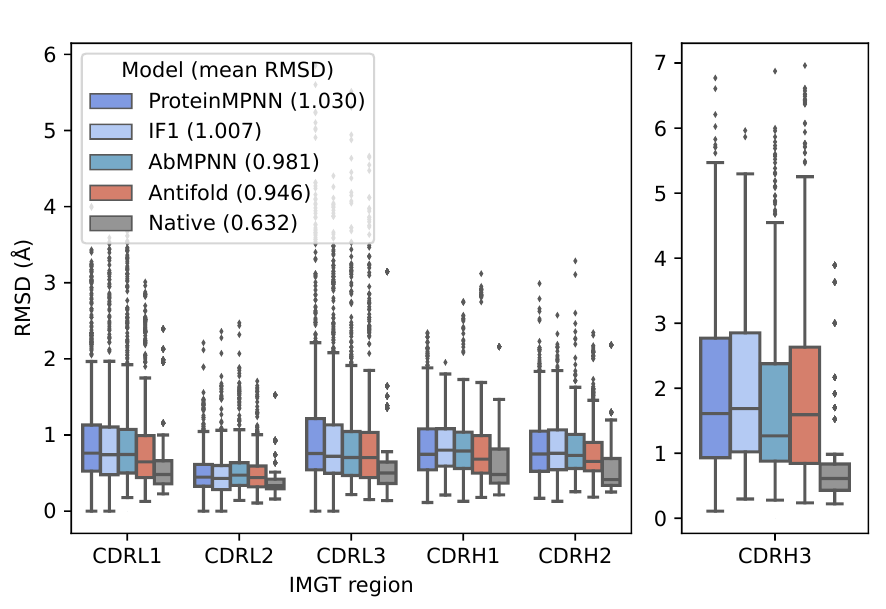}
  \caption{
  \textbf{AntiFold-designed sequences maintain the backbone structure.}
  CDR RMSDs between ABodyBuilder2-predicted and experimental structure backbones, for sequences sampled with ProteinMPNN, ESM-IF1, AntiFold and AbMPNN (temperature 0.20, see Supplementary Methods). CDR RMSDs between the AbodyBuilder2 model of the native sequence and the experimental structures are shown in gray. Mean CDR region RMSD values are shown in parentheses in the legend. We note that the RMSD calculation here only considers the modelling accuracy of the backbone and not side-chain atoms.}
  \label{fig:supp_a3_RMSD}
\end{figure*}

\begin{figure*}[htbp]
  \centering
  \includegraphics[width=0.70\textwidth]{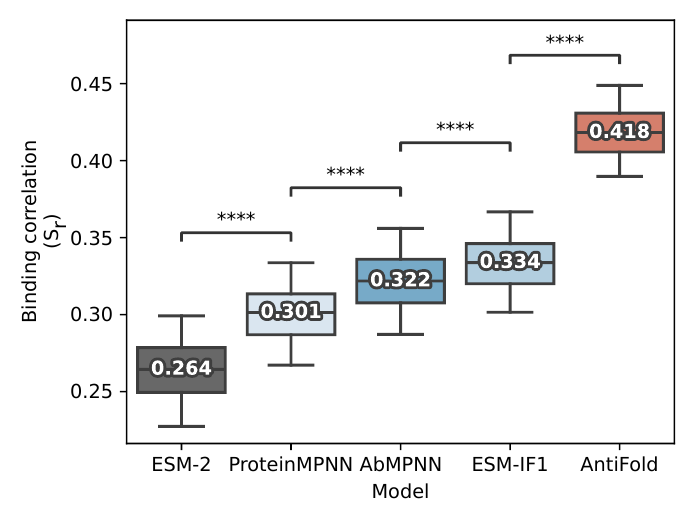}
  \caption{
  \textbf{Inverse folding probabilities correlate with antibody-antigen binding affinity.}
  Spearman's rank correlation between log-likelihood scores of the 2209 variants of the D44.1 anti-lysozyme antibody and the $log_{2}$ fold-change in binding affinity \cite{warszawski_vhvl_dms}. Error bars show 5-95th percentile range for the Spearman's rank correlation after bootstrapping 1000 times. The statistical significance from Mann-Whitney one-tailed U tests are shown (**** = p $<$ 0.00005).}
    \label{fig:antifold-fig3}
\end{figure*}
\setlength{\textfloatsep}{12pt}

\begin{figure*}[htb!]

  \centering
  \includegraphics[width=0.75\textwidth]{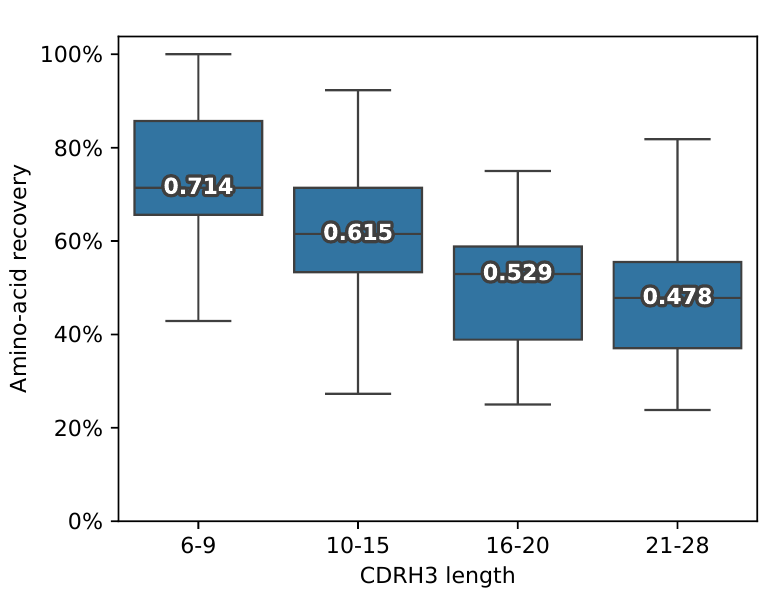}
  \caption{
  \textbf{AntiFold amino acid recovery is higher for shorter CDRH3 loops.} Test-set amino acid recovery by CDR3 heavy chain length.
}
  \label{fig:supp_a5_cdrlen}
\end{figure*}

\begin{figure*}[tb]
  \centering
  \includegraphics[width=0.75\textwidth]{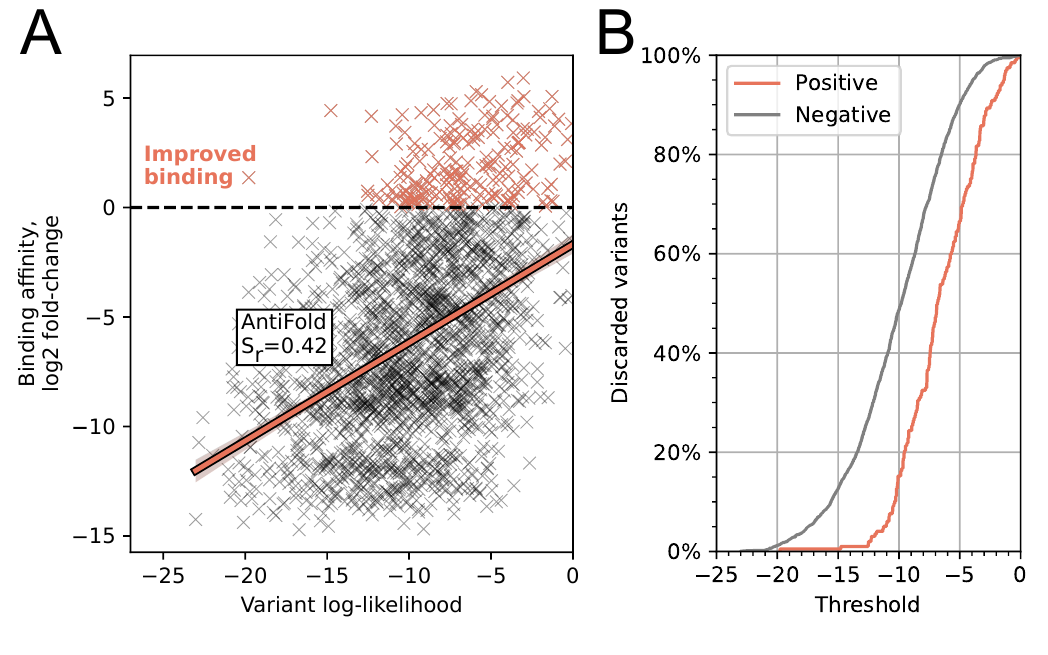}
  \caption{
  \textbf{AntiFold de-selection of lower binding affinity variants (anti-lysozyme deep mutational scan).}
A) Scatterplot of AntiFold variant log-likelihood versus experimental binding affinity values from \cite{warszawski_vhvl_dms}. Spearman's rank correlation and fitted ordinary least squares model shown. Variants with improved binding affinity ($log_{2}$ fold-change $>$ 0) are shown in orange (positives).
B) Cumulatively de-selected positive/negative variants at a given score threshold. At a minimum variant log-likelihood threshold of -11, ca 40\% of negatives and 5\% of positives were de-selected.
}
  \label{fig:supp_a4_dms}
\end{figure*}

\begin{figure*}[htbp]
  \centering
  \includegraphics[width=0.75\textwidth]{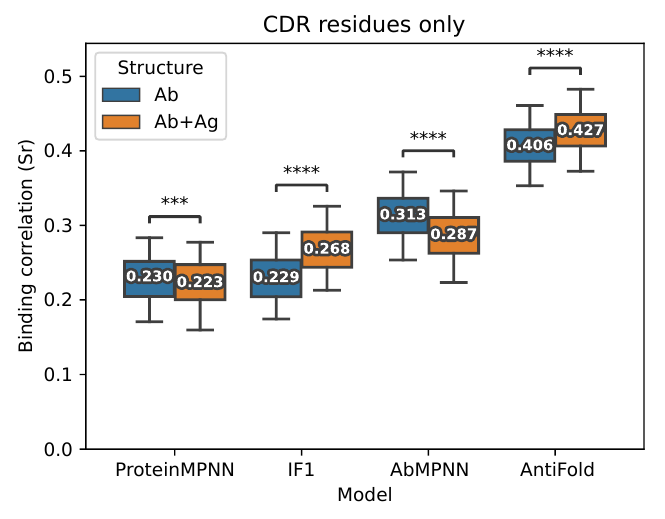}
  \caption{
  \textbf{AntiFold binding affinity prediction improves with antigen information.}
  Increase in Spearman's rank correlation between inverse folding model scores and experimental affinity values in the \cite{warszawski_vhvl_dms} deep mutational scan, excluding and including the antigen chain. Values bootstrapped 1000 times. Mann-Whitney one-tailed (less) U test shown (**** = p $<$ 0.00005).
}
  \label{fig:supp_a7_dms_abag}
\end{figure*}

\begin{figure*}[htbp]
  \centering
  \includegraphics[width=1.00\textwidth]{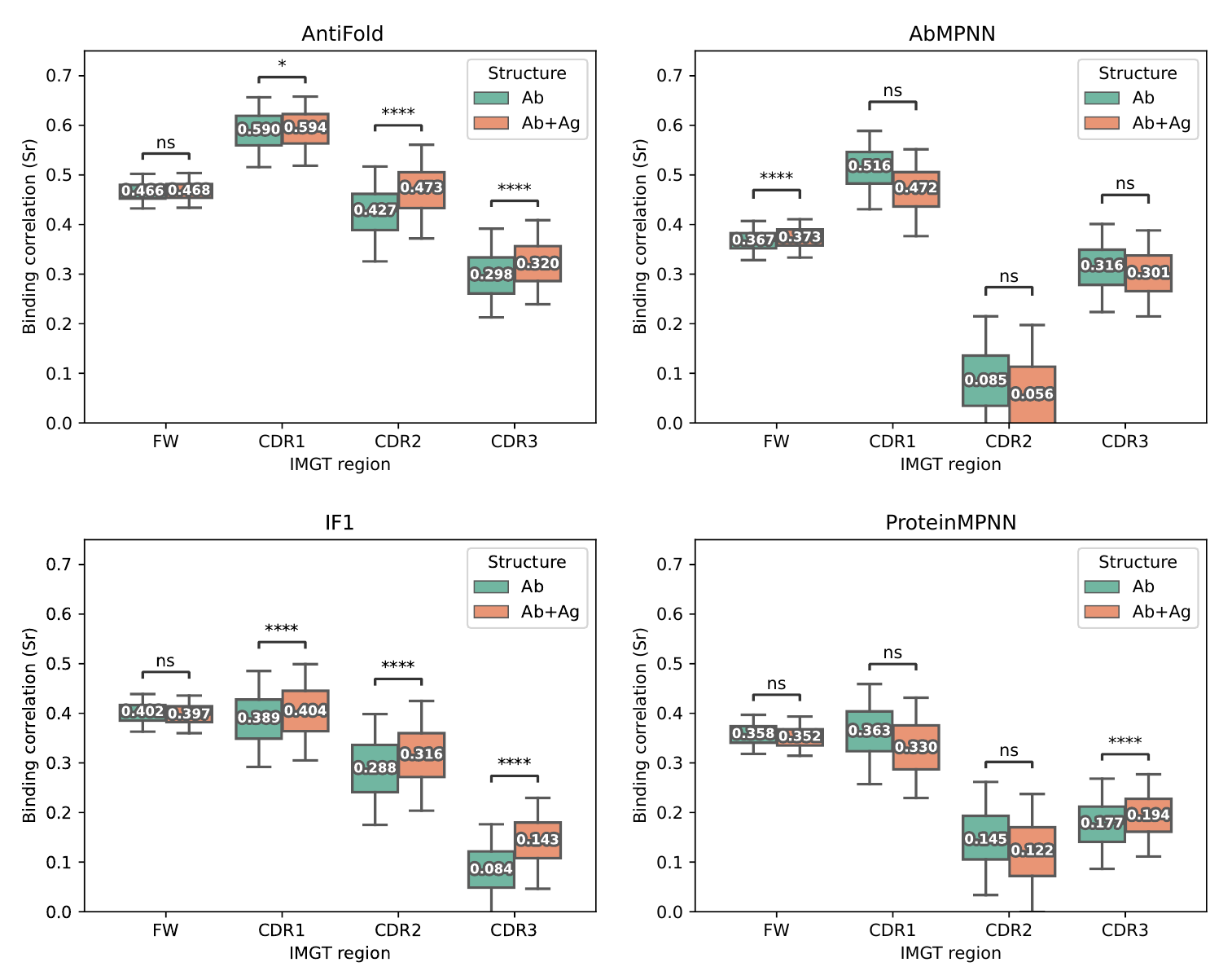}
  \caption{
  \textbf{Model antibody-antigen binding affinity prediction}
  Change in Spearman's rank correlation between inverse folding model scores and experimental affinity values in the \cite{warszawski_vhvl_dms} deep mutational scan, excluding and including the antigen chain and split by regions. Values bootstrapped 1000 times. Mann-Whitney one-tailed (less) U test shown (**** = p $<$ 0.00005).
}
  \label{fig:supp_a8_dms_abag}
\end{figure*}

\begin{figure*}[htbp]
  \centering
  \includegraphics[width=0.75\textwidth]{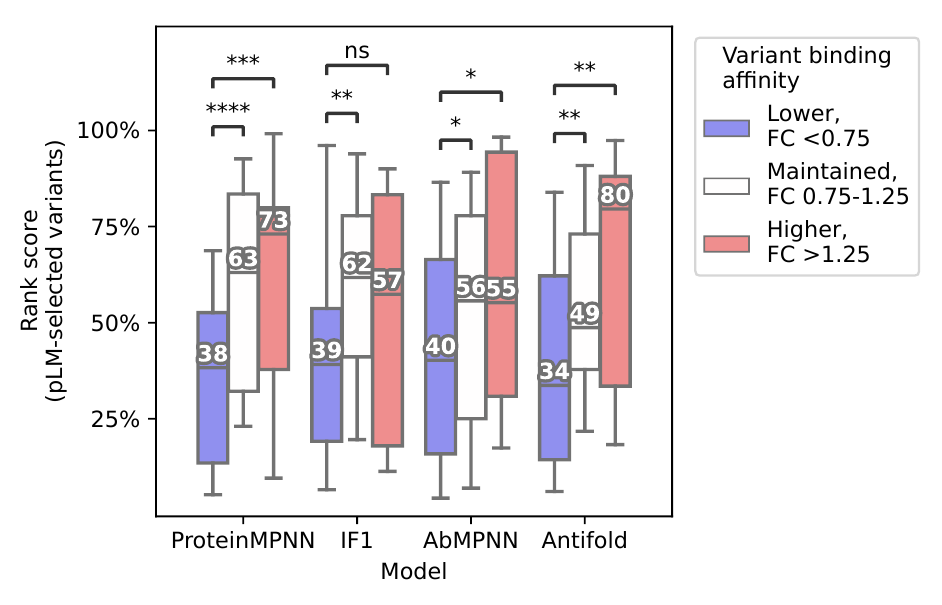}
  \caption{
  \textbf{AntiFold enables de-selection of protein language model-suggested variants.}
  Model rank-scoring performance across 124 variants of 7 antibodies from protein language model-guided affinity maturation experiments by \cite{hie_plm}. Variants are separated into lower (fold-change $<$0.75), maintained (fold-change 0.75-1.25) and higher (fold-change $>$1.25) binding affinity groups (see Supplementary Methods for details). Boxplot shows 5-95th percentile of variant scores (rank-normalized across the 124 variants), with median values and Mann-Whitney one-tailed (less) U test shown (** = p $<$ 0.005).}
    \label{fig:antifold-fig4}
\end{figure*}
\setlength{\textfloatsep}{12pt}

\begin{figure*}[htbp]
  \centering
  \includegraphics[width=0.75\textwidth]{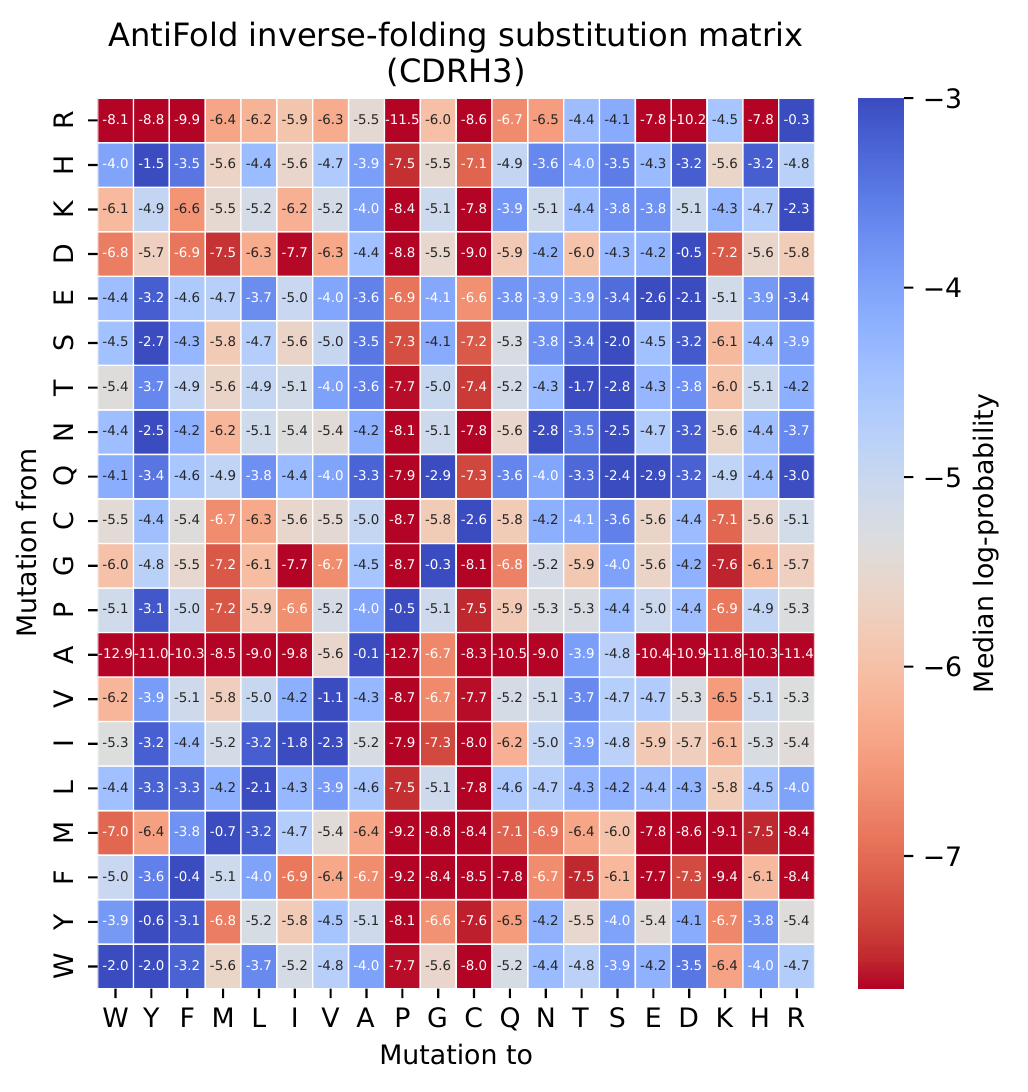}
  \caption{
  \textbf{AntiFold CDRH3 inverse folding probabilities capture physiochemical patterns.} Substitution matrix with median residue inverse folding probabilities (log) for mutations in the solved test-set CDR3 heavy chain regions.
}
  \label{fig:supp_a6_IFmatrix}
\end{figure*}

\newpage
\begin{landscape}
\begin{table}[h]
\centering
\caption{Fine-tuning parameter evaluation, applied to validation dataset (experimental, ``Exp", structures). The training (layer-wise learning rate decay, train masking) and testing (test masking) parameters are indicated. The values in the right side of the table represent amino acid recovery for a particular IMGT-region (FR: framework, CDR: complementarity-determining region). The highest value is shown in bold, the second-highest in italics.}
\label{tab:fine-tuning_exp}
\begin{tabular}{llll|lllllll}
\toprule
\textbf{Exp/Pred} & \textbf{Layer Decay} & \textbf{Train Masking} & \textbf{Test Masking} & \textbf{FR Avg.} & \textbf{CDRH1} & \textbf{CDRH2} & \textbf{CDRH3} & \textbf{CDRL1} & \textbf{CDRL2} & \textbf{CDRL3} \\
\midrule
Exp & -- & Shotgun & None & \textbf{0.845} & 0.695 & 0.606 & 0.532 & 0.597 & 0.584 & \textit{0.609} \\
Exp & -- & Span & None & 0.814 & 0.635 & 0.506 & 0.364 & 0.521 & 0.516 & 0.505 \\
Exp & -- & Shotgun + Span & None & \textit{0.842} & 0.675 & 0.601 & 0.525 & 0.570 & 0.559 & 0.582 \\
Exp & -- & Shotgun -- IMGT-Weighted& None & 0.835 & \textbf{0.708} & \textbf{0.640} & \textit{0.543} & \textbf{0.613} & \textbf{0.628} & \textbf{0.626} \\
Exp & -- & Span -- IMGT-Weighted& None & 0.807 & 0.636 & 0.511 & 0.365 & 0.535 & 0.521 & 0.516 \\
Exp & -- & Shotgun + Span -- IMGT-Weighted& None & 0.837 & 0.688 & 0.631 & 0.533 & 0.591 & 0.611 & 0.601 \\
Exp & \ding{51} & Shotgun & None & \textit{0.842} & \textbf{0.708} & 0.620 & \textit{0.543} & 0.601 & 0.567 & \textit{0.609} \\
Exp & \ding{51} & Span & None & 0.803 & 0.621 & 0.500 & 0.364 & 0.513 & 0.492 & 0.487 \\
Exp & \ding{51} & Shotgun + Span & None & 0.838 & 0.684 & 0.609 & 0.538 & 0.587 & 0.577 & 0.596 \\
Exp & \ding{51} & Shotgun -- IMGT-Weighted& None & 0.832 & \textbf{0.708} & \textit{0.636} & 0.541 & \textit{0.611} & \textit{0.614} & \textbf{0.626} \\
Exp & \ding{51} & Span -- IMGT-Weighted& None & 0.798 & 0.614 & 0.498 & 0.354 & 0.507 & 0.502 & 0.494 \\
Exp & \ding{51} & Shotgun + Span -- IMGT-Weighted& None & 0.833 & \textit{0.699} & 0.629 & \textbf{0.544} & 0.600 & 0.598 & 0.606 \\
\midrule
Exp & -- & Shotgun & CDRs & \textbf{0.832} & 0.520 & 0.388 & 0.310 & 0.439 & 0.438 & 0.437 \\
Exp & -- & Span & CDRs & 0.811 & \textit{0.622} & 0.507 & 0.348 & 0.521 & \textit{0.521} & 0.485 \\
Exp & -- & Shotgun + Span & CDRs & \textbf{0.832} & 0.587 & 0.477 & \textit{0.368} & 0.506 & 0.484 & 0.485 \\
Exp & -- & Shotgun -- IMGT-Weighted& CDRs & 0.827 & 0.608 & 0.496 & 0.343 & 0.520 & \textbf{0.545} & \textit{0.499} \\
Exp & -- & Span -- IMGT-Weighted& CDRs & 0.807 & \textbf{0.623} & \textit{0.512} & 0.354 & \textit{0.532} & 0.511 & \textbf{0.509} \\
Exp & -- & Shotgun + Span -- IMGT-Weighted& CDRs & \textit{0.828} & 0.604 & \textbf{0.532} & \textbf{0.380} & \textbf{0.541} & 0.511 & 0.493 \\
Exp & \ding{51} & Shotgun & CDRs & \textit{0.828} & 0.524 & 0.386 & 0.307 & 0.428 & 0.446 & 0.434 \\
Exp & \ding{51} & Span & CDRs & 0.800 & 0.599 & 0.483 & 0.330 & 0.494 & 0.467 & 0.470 \\
Exp & \ding{51} & Shotgun + Span & CDRs & 0.825 & 0.582 & 0.483 & 0.348 & 0.476 & 0.466 & 0.465 \\
Exp & \ding{51} & Shotgun -- IMGT-Weighted& CDRs & 0.824 & 0.580 & 0.476 & 0.350 & 0.478 & 0.498 & 0.466 \\
Exp & \ding{51} & Span -- IMGT-Weighted& CDRs & 0.795 & 0.606 & 0.508 & 0.343 & 0.490 & 0.479 & 0.485 \\
Exp & \ding{51} & Shotgun + Span -- IMGT-Weighted& CDRs & 0.822 & 0.609 & 0.497 & \textit{0.368} & 0.509 & 0.485 & 0.498 \\
\end{tabular}
\end{table}
\end{landscape}

\newpage
\begin{landscape}
\begin{table}[h]
\centering
\caption{Fine-tuning parameter evaluation, applied to validation dataset (predicted, ``Pred", structures). The training (layer decay, train masking) and testing (test masking) parameters are indicated. The values in the right side of the table represent amino acid recovery for a particular IMGT-region (FR: framework, CDR: complementarity-determining region). The highest value is shown in bold, the second-highest in italics.}
\label{tab:fine-tuning_pred}
\begin{tabular}{llll|lllllll}
\toprule
\textbf{Exp/Pred} & \textbf{Layer Decay} & \textbf{Train Masking} & \textbf{Test Masking} & \textbf{FR Avg.} & \textbf{CDRH1} & \textbf{CDRH2} & \textbf{CDRH3} & \textbf{CDRL1} & \textbf{CDRL2} & \textbf{CDRL3} \\
\midrule
Pred & -- & Shotgun & None & \textbf{0.856} & 0.703 & 0.617 & 0.519 & 0.600 & 0.611 & 0.604 \\
Pred & -- & Span & None & 0.816 & 0.639 & 0.505 & 0.373 & 0.531 & 0.506 & 0.499 \\
Pred & -- & Shotgun + Span & None & 0.851 & 0.697 & 0.602 & 0.510 & 0.580 & 0.563 & 0.596 \\
Pred & -- & Shotgun -- IMGT-Weighted& None & 0.850 & \textit{0.708} & \textit{0.640} & \textit{0.520} & \textbf{0.636} & \textit{0.625} & \textbf{0.635} \\
Pred & -- & Span -- IMGT-Weighted& None & 0.810 & 0.643 & 0.506 & 0.377 & 0.545 & 0.519 & 0.516 \\
Pred & -- & Shotgun + Span -- IMGT-Weighted& None & 0.844 & 0.701 & 0.628 & 0.513 & 0.589 & 0.604 & 0.602 \\
Pred & \ding{51} & Shotgun & None & \textit{0.853} & \textbf{0.710} & 0.626 & \textit{0.520} & 0.585 & 0.597 & 0.603 \\
Pred & \ding{51} & Span & None & 0.808 & 0.618 & 0.487 & 0.361 & 0.503 & 0.464 & 0.481 \\
Pred & \ding{51} & Shotgun + Span & None & 0.848 & 0.693 & 0.615 & 0.507 & 0.587 & 0.585 & 0.593 \\
Pred & \ding{51} & Shotgun -- IMGT-Weighted& None & 0.847 & 0.704 & \textbf{0.645} & \textbf{0.526} & \textit{0.620} & \textbf{0.632} & \textit{0.624} \\
Pred & \ding{51} & Span -- IMGT-Weighted& None & 0.803 & 0.615 & 0.509 & 0.359 & 0.512 & 0.499 & 0.493 \\
Pred & \ding{51} & Shotgun + Span -- IMGT-Weighted& None & 0.842 & 0.706 & 0.634 & 0.518 & 0.596 & 0.612 & 0.605 \\
\midrule
Pred & -- & Shotgun & CDRs & \textbf{0.844} & 0.535 & 0.395 & 0.327 & 0.438 & 0.444 & 0.444 \\
Pred & -- & Span & CDRs & 0.814 & 0.618 & 0.501 & 0.351 & 0.517 & 0.508 & 0.486 \\
Pred & -- & Shotgun + Span & CDRs & 0.840 & 0.603 & 0.481 & 0.374 & 0.517 & 0.473 & 0.478 \\
Pred & -- & Shotgun -- IMGT-Weighted& CDRs & \textit{0.841} & 0.622 & 0.504 & 0.356 & 0.522 & \textbf{0.534} & 0.492 \\
Pred & -- & Span -- IMGT-Weighted& CDRs & 0.810 & \textbf{0.630} & \textit{0.512} & 0.356 & \textit{0.536} & \textit{0.529} & \textit{0.499} \\
Pred & -- & Shotgun + Span -- IMGT-Weighted& CDRs & 0.836 & \textit{0.627} & \textbf{0.536} & \textbf{0.394} & \textbf{0.537} & 0.509 & \textbf{0.502} \\
Pred & \ding{51} & Shotgun & CDRs & 0.840 & 0.540 & 0.388 & 0.319 & 0.435 & 0.445 & 0.426 \\
Pred & \ding{51} & Span & CDRs & 0.805 & 0.600 & 0.488 & 0.341 & 0.498 & 0.481 & 0.464 \\
Pred & \ding{51} & Shotgun + Span & CDRs & 0.836 & 0.600 & 0.464 & 0.351 & 0.494 & 0.468 & 0.463 \\
Pred & \ding{51} & Shotgun -- IMGT-Weighted& CDRs & 0.837 & 0.586 & 0.476 & 0.361 & 0.482 & 0.498 & 0.475 \\
Pred & \ding{51} & Span -- IMGT-Weighted& CDRs & 0.800 & 0.610 & 0.503 & 0.344 & 0.493 & 0.496 & 0.487 \\
Pred & \ding{51} & Shotgun + Span -- IMGT-Weighted& CDRs & 0.835 & 0.612 & 0.487 & \textit{0.377} & 0.523 & 0.471 & 0.494 \\
\end{tabular}
\end{table}
\end{landscape}

\newpage
\begin{landscape}
\begin{table}[h]
\centering
\caption{Final model parameter evaluation, applied to validation dataset (experimental, ``Exp", and predicted, ``Pred", structures). Each model was trained with IMGT-weighted shotgun plus span masking for 1 epoch on the large predicted OAS structure dataset, followed by training on the experimental SAbDab dataset. The other training parameters (layer-wise learning rate decay and application of Gaussian noise to the predicted OAS structures) are indicated. The values in the right side of the table represent amino acid recovery for a particular IMGT-region (FR: framework, CDR: complementarity-determining region). The highest value is shown in bold, the second-highest in italics.}
\label{tab:fine-tuning_OAS}
\begin{tabular}{llll|lllllll}
\toprule
\textbf{Exp/Pred} & \textbf{Layer Decay} & \textbf{OAS Gaussian Noise} & \textbf{Test Masking} & \textbf{FR Avg.} & \textbf{CDRH1} & \textbf{CDRH2} & \textbf{CDRH3} & \textbf{CDRL1} & \textbf{CDRL2} & \textbf{CDRL3} \\
\midrule
Exp & - & - & None & \textbf{0.898} & 0.731 & \textbf{0.712} & 0.569 & \textbf{0.723} & \textit{0.736} & 0.718 \\
Exp & - & \ding{51} & None & \textbf{0.898} & \textit{0.735} & 0.698 & 0.566 & 0.716 & 0.702 & 0.713 \\
Exp & \ding{51} & - & None & \textit{0.895} & \textbf{0.741} & 0.700 & \textbf{0.584} & 0.716 & \textbf{0.741} & \textit{0.725} \\
Exp & \ding{51} & \ding{51} & None & 0.894 & 0.727 & \textit{0.702} & \textit{0.573} & \textit{0.720} & 0.728 & \textbf{0.727} \\
\midrule
Exp & - & - & CDRs & \textbf{0.894} & 0.680 & 0.637 & \textit{0.432} & \textit{0.677} & \textit{0.689} & \textbf{0.661} \\
Exp & - & \ding{51} & CDRs & \textbf{0.894} & \textbf{0.696} & 0.651 & \textbf{0.434} & \textbf{0.692} & 0.680 & \textit{0.659} \\
Exp & \ding{51} & - & CDRs & 0.890 & 0.675 & \textbf{0.657} & 0.431 & 0.666 & \textit{0.689} & 0.658 \\
Exp & \ding{51} & \ding{51} & CDRs & \textit{0.891} & \textit{0.681} & \textit{0.653} & 0.430 & 0.666 & \textbf{0.698} & 0.655 \\
\midrule
\midrule
Pred & - & - & None & \textbf{0.909} & \textbf{0.753} & \textit{0.716} & \textit{0.561} & 0.738 & 0.731 & \textit{0.722} \\
Pred & - & \ding{51} & None & 0.905 & 0.749 & 0.704 & 0.558 & 0.729 & 0.725 & \textit{0.722} \\
Pred & \ding{51} & - & None & \textit{0.907} & \textit{0.750} & \textbf{0.730} & \textbf{0.572} & \textbf{0.746} & \textbf{0.737} & \textbf{0.730} \\
Pred & \ding{51} & \ding{51} & None & 0.903 & 0.744 & 0.713 & 0.554 & \textit{0.744} & \textit{0.733} & 0.718 \\
\midrule
Pred & - & - & CDRs & \textbf{0.904} & \textit{0.706} & 0.650 & \textbf{0.445} & \textit{0.691} & \textit{0.687} & \textbf{0.665} \\
Pred & - & \ding{51} & CDRs & 0.901 & \textbf{0.709} & \textbf{0.657} & \textit{0.435} & \textbf{0.701} & \textbf{0.690} & \textit{0.658} \\
Pred & \ding{51} & - & CDRs & \textit{0.903} & 0.695 & \textit{0.654} & \textit{0.435} & 0.675 & 0.675 & 0.654 \\
Pred & \ding{51} & \ding{51} & CDRs & 0.898 & 0.699 & 0.647 & 0.433 & 0.682 & 0.682 & \textit{0.658} \\
\end{tabular}
\end{table}
\end{landscape}

\begin{landscape}
\begin{table}[h]
\centering
\caption{The model when developed without ESM-IF1 pretraining (i.e. with weights not initialised from ESM-IF1), with final parameters (layer-wise learning rate decay, IMGT-weighted shotgun plus span masking and application of Gaussian noise to the predicted OAS structures). The values in the right side of the table represent amino acid recovery for a particular IMGT-region (FR: framework, CDR: complementarity-determining region).}
\label{tab:fine-tuning_no_pre-training}
\begin{tabular}{ll|lllllll}
\toprule
\textbf{Exp/Pred} & \textbf{Test Masking} & \textbf{FR Avg.} & \textbf{CDR1H} & \textbf{CDR2H} & \textbf{CDRH3} & \textbf{CDR1L} & \textbf{CDR2L} & \textbf{CDR3L} \\
\midrule
Exp & None & 0.653 & 0.548 & 0.362 & 0.315 & 0.338 & 0.335 & 0.343 \\
\midrule
Exp & CDRs & 0.651 & 0.547 & 0.364 & 0.314 & 0.339 & 0.338 & 0.345 \\
\midrule
\midrule
Pred & None & 0.662 & 0.583 & 0.387 & 0.329 & 0.345 & 0.315 & 0.351 \\
\midrule
Pred & CDRs & 0.662 & 0.583 & 0.387 & 0.328 & 0.347 & 0.323 & 0.352 \\
\end{tabular}
\end{table}

\begin{table}[]
\centering
\caption{AntiFold performance in the case of an AlphaFold predicted antibody structure of median CDRH3 length ($\sim$15 residues).}
\label{tab:sanity1}
\begin{tabular}{@{}clll@{}}
\toprule
PDB                       & Structure     & \begin{tabular}[c]{@{}c@{}}Amino acid recovery\\ (\%)\end{tabular} & \begin{tabular}[c]{@{}c@{}}Correlation with Solved\\ ($S_{r}$)\end{tabular} \\ \midrule
\multirow{3}{*}{7M3N\_HL} & Solved        & 81.4\%                                                             & 1.000                                                                  \\ \cmidrule(l){2-4} 
                          & ABodyBuilder2 & 81.4\%                                                             & 0.941                                                                  \\ \cmidrule(l){2-4} 
                          & AlphaFold     & 81.4\%                                                             & 0.940                                                                  \\ \bottomrule
\end{tabular}
\end{table}

\end{landscape}

\end{document}